\newcounter{myctr}
\def\myitem{\refstepcounter{myctr}\bibfont\noindent\ifnum\themyctr>9\else\phantom{0}\fi\hangindent17pt\themyctr.\enskip}

\documentclass{ijqi-2e/ws-ijqi}

\usepackage{hyperref}
\usepackage[super,sort,compress]{cite}
\usepackage[dvipsnames]{xcolor}
\hypersetup{colorlinks=true, allcolors=blue, pdfusetitle=true}

\usepackage{zx-calculus}
\usepackage{quantikz}	

\usepackage{braket}

\makeatletter
\renewcommand\appendix{\refstepcounter{appendix}
	\setcounter{section}{0}%
        \setcounter{lemma}{0}
        \setcounter{theorem}{0}
	\setcounter{definition}{0}
        \setcounter{corollary}{0}
        \setcounter{conjecture}{0}        
	\setcounter{equation}{0}
	\@addtoreset{equation}{section}
	\renewcommand\theequation{\Alph{section}.\arabic{equation}}
	\renewcommand\sectionmark[1]{\markright{\thesection\ ##1:}} 
	\@ifstar\appendixstar\appendixnostar}
\newcommand\appendixnostar{
	\renewcommand\thesection{\appendixname\ \Alph{section}}}
\newcommand\appendixstar{
	\renewcommand\thesection{\appendixname}}
\makeatother

\begin{document}

\title{\uppercase{Three-qubit Deutsch-Jozsa in measurement-based quantum computing}}
\author{M. SCHWETZ and R. M. NOACK}
\address{Fachbereich Physik, Philipps Universität Marburg, Renthof 6\\
35037 Marburg, Germany\\
maximilian.schwetz@physik.uni-marburg.de}
\date{\today}

\maketitle

\begin{history}
\end{history}

\begin{abstract}
  Measurement-based quantum computing (MBQC), an alternate paradigm
  for formulating quantum algorithms, can lead to potentially more
  flexible and efficient implementations as well as to theoretical
  insights on the role of entanglement in a quantum algorithm.
  Using the graph-theoretical ZX-calculus, we describe and apply a general
  scheme for reformulating quantum circuits as MBQC implementations.
  After illustrating the method using the two-qubit Deutsch-Jozsa
  algorithm, we derive a ZX graph-diagram that encodes a general MBQC
  implementation for the three-qubit Deutsch-Jozsa algorithm. 
  This graph describes an 11-qubit cluster state on which single-qubit
  measurements are used to execute the algorithm.
  Particular sets of choices of the axes for the measurements can
  be used to implement any realization of the oracle.
  In addition, we derive an equivalent lattice cluster state for the algorithm.
\end{abstract}


\markboth{Schwetz and Noack}
{Three-qubit Deutsch-Jozsa in MBQC}

	
\section{Introduction}

Measurement-based quantum computing (MBQC) is an alternate
paradigm for quantum computing that is conceptually different than
the commonly-used circuit-based model \cite{MBQC-Review}.
As opposed to the circuit model, in which
entangling multi-qubit gates
effectuate the computing power of quantum mechanics,
MBQC pre-encodes the entanglement in a
highly entangled multi-qubit initial state, typically a cluster state,
that is prepared in advance.
Single-qubit measurements on the entangled initial state
are then all that is needed to unleash the
power of quantum computing in full generality
\cite{onewayQC, Cluster_QC, MBQC-Review}.
The MBQC model can lead to alternate experimental
implementations of quantum algorithms that can potentially be more
flexible and efficient than circuit-based implementations.
In addition, it is useful for exploring and understanding fundamental
aspects of quantum computing.

With regard to experimental utility,
since
the quantum-specific part of entanglement is
carried out
at the beginning when the initial cluster state is prepared,
it could be outsourced to a device that
is only responsible for generating entanglement.
This
prepared state
would then be sent
to a device that must only carry out
one-qubit measurements which are, in general, easy to do.
This device would, in effect, function as a universal computer.
On a fundamental level,
MBQC leads to a compelling alternate picture of quantum computing:
the power of quantum computing lies in the creation of a multi-qubit
entangled state, which is generally difficult.
The subsequent measurements can then be viewed as
classical tools
that utilize the previously
created resource of quantum entanglement for calculational 
purposes.
In this picture,
the ``quantumness'' and the classical operations are clearly separated.

Measurement-based quantum computing can be described
as a
sequence of measurements on the qubits, carried out along chosen axes,
of a cluster state.
Models such as
\emph{measurement calculus} \cite{Calculus}
or the \emph{monad} \cite{Monads} description
have been formulated to
try to encompass
the complexity of MBQC.
In this paper,
we will instead use the \emph{ZX-calculus}, a
graphical notation for graph states and graph-like operators
\cite{ZX-calculus}.
This intuitive and universal graphical calculus was introduced
for carrying out
derivations in multi-qubit quantum computation and information.
The diagrams are, on a low level, tensor-network descriptions of
quantum states and operators.
The ZX-calculus is also very well-suited to describe cluster states in
arbitrary geometries as well as to describe projective measurements.
Hence, it is a natural choice for
describing MBQC in an intuitive way.

The rules of the ZX-calculus are also capable
of translating an algorithm formulated in terms of
quantum circuits to the language of MBQC \cite{ZX-simplification}.
The rules are equations that tell the user how to transform
ZX-diagrams.
In the graph of a ZX-diagram, they range from single-node identities
to large-scale node and edge manipulations.
One use for these rules is
to simplify diagrams and thus to
optimize quantum circuits.
However, in this work, we use the ZX-calculus
to translate quantum circuits to the language
of ZX-diagrams,
which we then refine to reformulate the algorithm as a MBQC calculation.
As a concrete and, hopefully, intuitive, example,
we will formulate
a measurement-based
implementation
of the Deutsch-Jozsa algorithm for
three qubits.
As an introduction, we will also describe the Deutsch-Jozsa algorithm
for one and two qubits.
For the two-qubit case,
a measurement-based
implementation can already be found in the literature
\cite{4qubit-DJ,Cluster-Deutsch-Jozsa}.

The \emph{Deutsch-Jozsa-Algorithm} was one of the first quantum
algorithms to be
proven to be computationally superior to
functionally equivalent algorithms on classical computers; in fact,
it was specifically designed to show the power of quantum computing.
A  one-qubit precursor to
the algorithm was originally proposed by
D.~Deutsch \cite{Deutsch};
however, this original variant is not deterministic.
The algorithm was subsequently extended to treat $n$ qubits,  
to be deterministic, and to be less
demanding computationally by Deutsch and Jozsa \cite{Deutsch-Jozsa}
and by Cleve \emph{et al.} \cite{Algorithms-revisited}.
In the present day,
the Deutsch-Jozsa-Algorithm is typically used
as a useful introduction
to \emph{quantum advantage}, the theoretical
superiority of quantum computers over classical
computers.

The aim of the algorithm is
to determine if a function is \emph{constant} or
\emph{balanced}.
We take
a function $f: Z_2^n \ni \boldsymbol{\sigma} \rightarrow
f(\boldsymbol{\sigma}) \in Z_2$ that maps a binary representation $\boldsymbol{\sigma}$ of
length $n$ to one bit $f(\boldsymbol{\sigma})$.
We call $f(\boldsymbol{\sigma})$ \emph{constant} if the outcome is the same for
all choices of $\boldsymbol{\sigma}$, i.e.,
$\forall \boldsymbol{\sigma} \in Z_2^n: f(\boldsymbol{\sigma}) = 0$
or $\forall \boldsymbol{\sigma} \in Z_2^n: f(\boldsymbol{\sigma}) = 1$.
We call the function
\emph{balanced} if exactly one-half of the outcomes are $0$ and
one-half are $1$,
i.e.,  $|\{\boldsymbol{\sigma} \in Z_2^n: f(\boldsymbol{\sigma}) = 0\}|
= |\{\boldsymbol{\sigma} \in Z_2^n: f(\boldsymbol{\sigma}) = 1\}|$.
It is important to emphasize
that the algorithm is based on the
premise
that $f$ can only be constant or balanced and
nothing else.
The possible variants of $f$ are tabulated
in Table~\ref{tab:function1}.

It is easy to show
that a quantum computer can solve this problem with less
work than a classical computer.
Suppose we have an oracle that returns the bit $f(\boldsymbol{\sigma})$ for
any $\boldsymbol{\sigma}$ that we feed into it.
A classical computer requires $2^{n-1} + 1$ queries of the oracle
to solve the problem in the worst case, in which
$f$ is balanced, but
the first $2^n/2 = 2^{n-1}$ queries all yield the same output.
On a quantum device, one can determine the character of the function
$f$ with just one call to the oracle, i.e., at
constant cost.

\begin{table}
\begin{center}
\caption{\label{tab:function1}Output variants of a one-bit boolean function.
Variants ($i$) and ($ii$) are constant, variants ($iii$) and ($iv$) are balanced.}
\begin{tabular}{ccccc}
\hline\hline
 & \multicolumn{4}{c}{Variants} \\
 & ($i$) & ($ii$) & ($iii$) & ($iv$) \\
 \hline
 $f(0)$ & 0 & 1 & 0 & 1 \\
 $f(1)$ & 0 & 1 & 1 & 0 \\
 \hline\hline
\end{tabular}
\end{center}
\end{table}

The canonical way of formulating
the Deutsch-Jozsa algorithm makes use of
an auxiliary qubit.
Here, however,
we will work with
a variant of the algorithm that uses no auxiliary qubit, but only the
query qubits.
In the following,
we will profit from the reduced
number of qubits as we translate the algorithm to measurement-based
quantum computing.

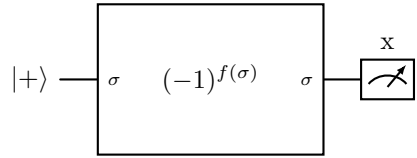
\begin{figure}
\begin{center}
\begin{quantikz}
 \lstick{$\ket{+}$} & \gate[][3cm][2cm]{\;(-1)^{f(\sigma)}\;} \gateinput{$\sigma$}\gateoutput{$\sigma$} & \meter{$x$}
\end{quantikz}
\caption{\label{fig:DJ-algorithm1}Deutsch-Jozsa algorithm for one qubit.
 The oracle adds a phase of $-1$ to each basis element
 $\ket{\sigma}$ when $f(\sigma) = 1$.
 Measurement of $\ket{+}$
 reveals
 that $f$ is constant, $\ket{-1}$
 that $f$ is balanced.}
\end{center}
\end{figure}

The single-qubit-input variant of the one-bit Deutsch-Jozsa algorithm
is shown in
Fig.~\ref{fig:DJ-algorithm1}.
(All quantum circuits were typeset with the help of the
\emph{Quantikz}-package \cite{Quantikz}.)
We prepare one working qubit in the state
$\ket{+}= \ket{0}+\ket{1}$
(omitting normalization here and throughout this work).
We then apply an oracle operation to the working qubit, where
the effect of the oracle depends on
the character of the instance of the function $f$ under
investigation.
If $f(\sigma)$ is 1, then the oracle
adds a phase of $-1$ to the projection
onto the basis state $\ket{\sigma}$.
That is, it transforms the input state as
\begin{equation}
  \ket{+}= \ket{0}+ \ket{1} \
  \xrightarrow{\text{oracle}} \
  (-1)^{f(0)} \ket{0}+ (-1)^{f(1)}
  \ket{1} \, . \label{eq:oracle1} 
\end{equation}
The oracle can be realized by the identity operation
if $f$ is constant
and by a Pauli-$Z$-gate if $f$ is balanced.

After the application of the oracle, the only remaining step
is to measure the working qubit in the $x$-basis.
If the measurement yields the outcome corresponding to the basis state
$\ket{+}$, then $f$ is constant.
For the other outcome, $f$ is balanced.
In the one-qubit case, it can easily be checked that the oracle of
Eq.~(\ref{eq:oracle1}) will transform $\ket{+}$ to $\pm \ket{+}$ if
and only if $f(0) = f(1)$.
Note that the overall phase of $\pm 1$
does not have physical relevance
and cannot be measured.
If, on the other hand,
$f(0) \ne f(1)$, the oracle will transform
$\ket{+}$ to $\pm \ket{-}$.
Hence, a measurement in the
$x$-basis will determine the character of $f$.

Our aim here is to derive the three-qubit measurement-based description of
the Deutsch-Jozsa algorithm in a systematic way using the ZX-calculus.
In order to introduce the ZX-calculus and give an illustration of
how to translate an algorithm represented as a quantum circuit to MBQC, 
we will start with
the non-canonical two-qubit Deutsch-Jozsa algorithm, i.e., the variant
without an auxiliary qubit in Sec.~\ref{sec:circuit2}.
We will reformulate
the two-qubit algorithm in terms of MBQC in
Sec.~\ref{sec:ZX2}.
After having illustrated the concepts and procedure for the two-qubit
case, 
we will go on to describe the quantum-circuit implementation of the
three-qubit algorithm 
in Sec.~\ref{sec:circuit3}.
We will observe
that the algorithm gains a new level of complexity for
more than two qubits.
Using the ZX-calculus, we will then reformulate the algorithm in terms
of MBQC in Sec.~\ref{sec:ZX3}.
The result will be a special geometric graph cluster state and a
recipe for choosing a set of 
measurements to execute the Deutsch-Jozsa algorithm for any instance
of the oracle.
We will also provide a ZX-diagram for implementing the algorithm as
a rectangular cluster-state lattice.
Finally, in Sec.~\ref{sec:Conclusion}, we will discuss
what we have learned from
this work
and give an outlook for potential future
work.


\section{Two-qubit Deutsch-Jozsa}
\subsection{Circuit model}
\label{sec:circuit2}

The Deutsch-Jozsa algorithm for two or more qubits
is a relatively straightforward extension of the one-qubit case.
Here we will treat the two-qubit case explicitly, taking it as an example to illustrate
how to use ZX-calculus to systemically convert the circuit-model implementation
to a measurement-based implementation.
A description of the original measurement-based implementation of the 
two-qubit Deutsch-Jozsa algorithm can be found in Ref.~\citen{Cluster-Deutsch-Jozsa}.
We remark that Ref.~\citen{Cluster-Deutsch-Jozsa} uses the variant of the algorithm
that utilizes an auxiliary qubit.
This variant requires entangling operations.
In the following, we will show, in a visually evident way using ZX-diagrams,
that no entanglement is necessary if one utilizes
the formulation of the Deutsch-Jozsa algorithm with no auxiliary qubit.

For the two-bit case, the function $f(\sigma_0, \sigma_1)$ has
eight possible realizations,
each of which must be either constant or balanced;
they are listed in Table~\ref{tab:function2}.
The quantum algorithm depicted in Fig.~\ref{fig:2bit-DJ} in
schematic circuit representation, is used to determine which
character
$f$ has.
The two working qubits are prepared in the $\ket{+}$-state.
The oracle then adds a phase of $-1$ to all computational basis states
$\ket{\sigma_0, \sigma_1} \; (\sigma_{0/1} \in \{0, 1\})$ if and only if
$f(\sigma_0, \sigma_1)= 1$.
A measurement of both qubits in the $x$-basis reads out
the character
of $f$:
if $\ket{++}$ is measured, then $f$ is constant; if any other state is
measured, then $f$ is balanced.

\begin{table}
\begin{center}
\caption{\label{tab:function2}Outputs of a two-bit boolean function
  that is either \emph{constant} or \emph{balanced}.
  Variants ($i$) and ($ii$) are constant, while variants ($iii$)
  through ($viii$) are balanced.
  The last four rows show the control angles for the measurement-based
  version of the algorithm, which is introduced in
  diagram~(\ref{eq:raw2bit_oracle}).
}
\begin{tabular}{ccccccccc}
 \hline\hline
 & \multicolumn{8}{c}{Variants} \\
 & ($i$) & ($ii$) & ($iii$) & ($iv$) & ($v$) & ($vi$) & ($vii$) & ($viii$) \\
 \hline
 $f(0, 0)$ & 0 & 1 & 0 & 0 & 0 & 1 & 1 & 1 \\
 $f(0, 1)$ & 0 & 1 & 0 & 1 & 1 & 0 & 0 & 1 \\
 $f(1, 0)$ & 0 & 1 & 1 & 0 & 1 & 0 & 1 & 0 \\
 $f(1, 1)$ & 0 & 1 & 1 & 1 & 0 & 1 & 0 & 0 \\
 $\alpha_0$ & 0 & 0 & 0 & 0 & $\pi$ & $\pi$ & 0 & $\pi$ \\
 $\alpha_1$ & 0 & 0 & $\pi$ & 0 & 0 & $\pi$ & 0 & $\pi$ \\
 $\alpha_2$ & 0 & 0 & 0 & $\pi$ & $\pi$ & $\pi$ & $\pi$ & 0 \\
 $\alpha_3$ & 0 & 0 & 0 & 0 & 0 & $\pi$ & $\pi$ & 0 \\
 \hline\hline
\end{tabular}
\end{center}
\end{table}

\begin{figure}
	\centering
	\begin{quantikz}[row sep={1cm,between origins}]
		\lstick{$\ket{+}$} & \gate[wires=2, label style={yshift=0cm}, disable auto height][3cm][1cm]{(-1)^{f(\sigma_0, \sigma_1)}} \gateinput{$\sigma_0$}\gateoutput{$\sigma_0$} & \meter{x} \\
		\lstick{$\ket{+}$} & \qw \gateinput{$\sigma_1$}\gateoutput{$\sigma_1$}& \meter{x}
	\end{quantikz}
	\caption{Two-qubit Deutsch-Jozsa algorithm.
         The variants of the oracle are depicted in Fig.~\ref{fig:DJ-variants}.}
	\label{fig:2bit-DJ}
\end{figure}
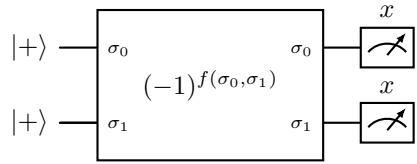

\begin{figure}
\begin{align*}
	(i) \quad &\begin{quantikz}[align equals at=1.5, row sep={0.3cm}, ampersand replacement=\&]
		\&\hphantomgate{Z} \gategroup[wires=2, steps=1,style={inner sep=2pt}]{} \& \qw \\
		\qw \& \hphantomgate{Z} \& \qw
	\end{quantikz}
	& (ii) \quad & \begin{quantikz}[align equals at=1.5, row sep={0.3cm}, ampersand replacement=\&]
		\&\hphantomgate{Z} \gategroup[wires=2, steps=1,style={inner sep=2pt}]{} \& \qw \\
		\qw \& \hphantomgate{Z} \& \qw
	\end{quantikz} \\
	(iii) \quad &\begin{quantikz}[align equals at=1.5, row sep={0.3cm}, ampersand replacement=\&]
	\& \gate{Z} \gategroup[wires=2, steps=1,style={inner sep=2pt}]{}\& \qw \\
		\qw \& \hphantomgate{Z} \& \qw
	\end{quantikz}
	& (iv) \quad & \begin{quantikz}[align equals at=1.5, row sep={0.3cm}, ampersand replacement=\&]
	\& \hphantomgate{Z} \gategroup[wires=2, steps=1,style={inner sep=2pt}]{}\& \qw \\
		\qw \& \gate{Z} \& \qw
	\end{quantikz} \\
	(v) \quad &\begin{quantikz}[align equals at=1.5, row sep={0.3cm}, ampersand replacement=\&]
	\& \gate{Z} \gategroup[wires=2, steps=1,style={inner sep=2pt}]{}\& \qw \\
		\qw \& \gate{Z} \& \qw
	\end{quantikz}
	& (vi) \quad & \begin{quantikz}[align equals at=1.5, row sep={0.3cm}, ampersand replacement=\&]
	\& \gate{Y} \gategroup[wires=2, steps=1,style={inner sep=2pt}]{}\& \qw \\
		\qw \& \gate{Y} \& \qw
	\end{quantikz} \\
	(vii) \quad &\begin{quantikz}[align equals at=1.5, row sep={0.3cm}, ampersand replacement=\&]
	\& \hphantomgate{Y} \gategroup[wires=2, steps=1,style={inner sep=2pt}]{}\& \qw \\
		\qw \& \gate{Y} \& \qw
	\end{quantikz}
	& (viii) \quad & \begin{quantikz}[align equals at=1.5, row sep={0.3cm}, ampersand replacement=\&]
	\& \gate{Y} \gategroup[wires=2, steps=1,style={inner sep=2pt}]{}\& \qw \\
		\qw \& \hphantomgate{Y} \& \qw
	\end{quantikz}
\end{align*}
\caption{
  Circuit-model implementations of all variants of the oracle for
  the Deutsch-Jozsa algorithm on two input qubits as tabulated in Table \ref{tab:function2}.
  Here $X$ and $Y$ are Pauli gates.
  Overall phases, which do not alter the outcome of measurements, are ignored.}
	\label{fig:DJ-variants}
\end{figure}
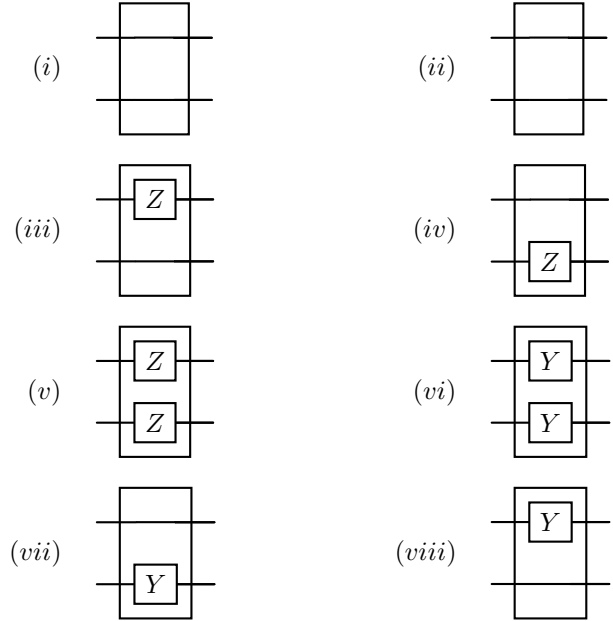

Within
a circuit model without auxiliary qubits,
the eight variants of the
oracle can be implemented as shown in Fig.~\ref{fig:DJ-variants}.
Circuits for the variant
with an auxiliary qubit can be found, for example, in
Ref.~\citen{Cluster-Deutsch-Jozsa}.
The oracle will apply one of these circuits to the two working
qubits.
Note that information about the character of $f$ is hidden to us; 
it is contained in the choice of the circuit used by the oracle.
We know only that the circuit is restricted to
one of the eight variants depicted in Fig.~\ref{fig:DJ-variants}.


\subsection{Measurement-based model}
\label{sec:ZX2}

We now turn
from the circuit model
to the measurement-based model.
In order
to systematically formulate
the Deutsch-Jozsa algorithm in the
measurement-based scheme,
we must rewrite
the oracle modes as measurement patterns on
a cluster state.
To switch to the language of measurement-based quantum computing, we
use the
ZX-calculus~\cite{ZX-calculus}.
This intuitive and universal graphical calculus was introduced
to carry out
derivations in multi-qubit quantum computation and information.
Here we will give a short summary of its features
and will introduce only
the rules that are needed for
this work.
For a comprehensive overview describing the formalism and illustrating
the power
of ZX-calculus,
we refer the reader to
Ref.~\citen{Circuit-extraction-tale}.

The ZX-calculus is a
description of quantum states or quantum
operations in terms of graphs, which link two types of nodes:
$Z$-nodes, colored green here (or lightly shaded for the visually impaired
reader)
and $X$-nodes, colored red (darkly shaded) here.
They are, on a tensor level, defined as
\begin{equation}
	\zx{\leftManyDots{} \zxZ{\alpha} \rightManyDots{}} = \ket{0\dotsc 0}\bra{0\dotsc0} + \mathrm{e}^{\mathrm{i}\alpha} \ket{1\dotsc1}\bra{1\dotsc1}
\end{equation}
and
\begin{equation}
	\zx{\leftManyDots{} \zxX{\alpha} \rightManyDots{}} = \ket{+\dotsm +}\bra{+\dotsm+} + \mathrm{e}^{\mathrm{i}\alpha} \ket{-\dotsm-}\bra{-\dotsm-} \; ,
\end{equation}
where $\ket{\pm}=\ket{0}\pm\ket{1}$
(as mentioned above,  we neglect normalization factors).
Incoming and outgoing legs represent degrees of freedom, i.e., tensor
indices.
We note that all ZX-diagrams were typeset with the \emph{zx-calculus}
package \cite{ZX-TeX}.

Connecting nodes via edges leads to diagrams that represent either quantum
states or quantum operators, depending on the configuration of the outgoing edges.
As both states and operators are represented as tensors, we will
not explicitly distinguish between
them; the nature of a diagram should be clear from the context.
As a simple example, the eigenstate
of the $x$-Pauli matrix can be
written as $\zx{\zxZ{} \rar &[\zxwCol] \zxN{}}= \ket{+}$, where we
omit the relative phase angle $\alpha$
of the node when it is zero.

Calculations are carried out in the ZX-calculus by applying rules to
transform diagrams.
Rules can be derived by expressing a diagram in tensor notation and
transforming, typically simplifying, the tensor contractions.
A comprehensive list of
rules is given in Ref.~\citen{Circuit-extraction-tale}.
A commonly used special simplification is the representation of the
Hadamard gate as a single blue
dashed line:
\begin{equation}
 H = \zx{\zxN{} &[\zxwCol] \zxFracZ{\pi}{2} \lar\rar &
 \zxFracX{\pi}{2} \rar & \zxFracZ{\pi}{2} \rar &[\zxwCol]
 \zxN{}} = \zx{\zxN{} \ar[r, blue, dashed] &[2em]
 \zxN{}} \  .\label{eq:rule_Hadamard}
\end{equation}
In this work, we will utilize the following subset of ZX-calculus rules:
\begin{align}
	\begin{ZX}[ampersand replacement=\&]
		\zxN{} \&[\zxwCol]\&[\zxwCol] \zxN{} \\
		\zxN{} \& \zxX{\alpha} \ar[l, 3 vdots] \ar[lu, (] \ar[dl, )] \ar[ur, )] \ar[dr, (] \ar[r, 3 vdots] \& \zxN{} \zxN{} \& \\
		\zxN{} \&\& \zxN{} 
	\end{ZX} &=
	\begin{ZX}[ampersand replacement=\&]
		\zxN{} \&[\zxwCol]\&[\zxwCol] \zxN{} \\
		\zxN{} \& \zxZ{\alpha} \ar[l, 3 vdots] \ar[lu, blue, dashed, (] \ar[dl, blue, dashed, )] \ar[ur, blue, dashed, )] \ar[dr, blue, dashed, (] \ar[r, 3 vdots] \& \zxN{} \\
		\zxN{} \&\& \zxN{}
	\end{ZX} \; , \label{eq:rule_colorswap} \\
	\begin{ZX}[ampersand replacement=\&]
		\leftManyDots{} \zxZ{\alpha} \ar[d, 3 dots]\ar[d, (]\ar[d, )] \rightManyDots{} \\[\zxWRow]
		\leftManyDots{} \zxZ{\beta} \rightManyDots{}
	\end{ZX} &=
	\begin{ZX}[ampersand replacement=\&]
		\leftManyDots{} \zxZ{\alpha+\beta} \rightManyDots{}
	\end{ZX} \; , \label{eq:rule_contraction}\\
	\begin{ZX}[ampersand replacement=\&]
		\zxN{} \&[\zxwCol] \zxZ{} \ar[l, blue, dashed] \ar[r, blue, dashed] \&[\zxwCol] \zxN{}
	\end{ZX} &=
	\begin{ZX}[ampersand replacement=\&]
		\zxN{} \rar \&[\zxWCol] \zxN{}
	\end{ZX} \; , \label{eq:rule_hadamard} \\
	\begin{ZX}[ampersand replacement=\&]
		\zxN{} \& \&[\zxWCol] \zxN{} \\
		\zxX{} \rar \& \zxZ{\alpha} \ar[ur, )] \ar[dr, (] \ar[r, 3 vdots] \& \zxN{} \\
		\zxN{} \& \& \zxN{}
	\end{ZX} &=
	\begin{ZX}[ampersand replacement=\&]
		\zxX{} \rar \&[\zxWCol] \zxN{} \\
		\zxN{} \ar[r, 3 vdots] \& \zxN{} \\
		\zxX{} \rar \& \zxN{}
	\end{ZX} \; . \label{eq:rule_decoupling}
\end{align}
Note that rules (\ref{eq:rule_colorswap})-(\ref{eq:rule_decoupling})
are also valid when Z-nodes and X-nodes are interchanged in the
diagrams.

For our purposes, the important feature of ZX-diagrams is that they
can be naturally used to describe
measurement-based quantum computing.
In order to see how to do this, 
we realize
that if we have a ZX-diagram with only Z-nodes
(green), and all edges are Hadamard edges (blue-dashed), then the state
represented by the diagram is a \emph{cluster state} in which
the Z-nodes represent the qubits \cite{Circuit-extraction-tale}.
A projective measurement of a qubit along an axis in the $x$-$y$-plane
with outcome 0 can be written as the projector
$\zx{\zxN{} &[\zxWCol] \zxZ{\alpha} \lar}$, where $\alpha$ is the
angle of the basis in the plane.
For example, a measurement along
the $x$-axis with outcome
0, corresponding
to the projected state $\ket{+}=\ket{0}+\ket{1}$, would be
represented by $\zx{\zxN{} &[\zxWCol] \zxZ{} \lar}$.
Since
all nodes in a cluster state are Z-nodes, one can contract this
measurement projection to the respective qubit using
rule (\ref{eq:rule_contraction}).
Hence, a ZX-diagram consisting of only Z-nodes and Hadamard edges
can also be understood as a sequence of measurements on a cluster
state.
On a technical note,
in ZX-calculus, it is usually assumed for simplicity
that any
measurement always has the
outcome 0.
If the other outcome, 1, were to occur instead,
this ``error'' could be propagated
through the network so that either following measurement axes need to
be adjusted or the final measurement outcome has to be reversed.
We will not elucidate this procedure further here;
see Ref.~\citen{Circuit-extraction-tale} for a detailed explanation.

For the circuits in Fig.~\ref{fig:DJ-variants},
we need to translate the Pauli $Z$ and $Y$ gates.
It is evident
that
$Z = \zx{\zxN{} &[\zxWCol] \zxX{\pi} \lar\rar &[\zxWCol] \zxN{}}$
and
$Y \propto XZ = \zx{\zxN{} &[\zxWCol] \zxX{\pi} \lar\rar & \zxZ{\pi}
  \rar &[\zxWCol] \zxN{}}$.
Using these translations,
we can write
a general instance of the two-bit oracle as
\begin{equation}\begin{ZX}
	\zxN{} &[\zxwCol] \zxX{\alpha_0} \lar\rar & \zxZ{\alpha_1} \rar &[\zxwCol] \zxN{} \\
	\zxN{} & \zxX{\alpha_2} \lar\rar & \zxZ{\alpha_3} \rar & \zxN{}
\end{ZX} \; ,
  \label{eq:raw2bit_oracle}
\end{equation}
where the angles $\alpha_i$ are chosen by the oracle according to
Table~\ref{tab:function2} to represent the different combinations of $Y$
and $Z$.
Using the Hadamard abbreviation (\ref{eq:rule_Hadamard}), we can translate
diagram (\ref{eq:raw2bit_oracle})
to the measurement-based quantum-computing description
\begin{equation}\begin{ZX}
	\zxN{} &[\zxwCol] \zxZ{} \lar \ar[r, blue, dashed] &[\zxHCol] \zxZ{\alpha_0} \ar[r, blue, dashed] &[\zxHCol] \zxZ{\alpha_1} \rar &[\zxwCol] \zxN{} \\
	\zxN{} & \zxZ{} \lar\ar[r, blue, dashed] & \zxZ{\alpha_2} \ar[r, blue, dashed] & \zxZ{\alpha_3} \rar & \zxN{}
\end{ZX} \; . \label{eq:2-qubit-ZX-oracle} \end{equation}
Note that we have added an extra identity $\zx{\zxN{} &[\zxwCol]
  \zxZ{} \lar\rar &[\zxwCol] \zxN{}} = \zx{\zxN{} &[\zxWCol]
  \zxN{} \lar}$ to the left of each row so that the Hadamard edge does
not protrude.

Recall that the Deutsch-Jozsa algorithm starts with the working qubits
in the $\ket{+}$ state and ends with a measurement $\bra{+}$ of the
same state.
In the ZX-calculus, both are represented by $\zx{\zxN{}&\zxZ{}\lar}$.
By adding this to diagram (\ref{eq:2-qubit-ZX-oracle}) and using rule (\ref{eq:rule_contraction}) in the trivial way, $\zx{\zxZ{} \rar & \zxZ{} \rar &[\zxwCol] \zxN{}} = \zx{\zxZ{} \rar &[\zxwCol] \zxN{}}$ we obtain
the diagram for
the full algorithm:
\begin{equation}\begin{ZX}
	\zxZ{} \ar[r, blue, dashed] &[\zxHCol] \zxZ{\alpha_0} \ar[r, blue, dashed] &[\zxHCol] \zxZ{\alpha_1} \\
	\zxZ{} \ar[r, blue, dashed] &[\zxHCol] \zxZ{\alpha_2} \ar[r, blue, dashed] &[\zxHCol] \zxZ{\alpha_3}
  \end{ZX} \; . \label{eq:2-qubit-ZX-algorithm} \end{equation}
According to our description above, diagram
(\ref{eq:2-qubit-ZX-algorithm}) can be interpreted as two
linear cluster states, each with three qubits as a resource state.
The qubits of the upper linear cluster state are measured along
axes 0, $\alpha_0$, and $\alpha_1$, respectively.
The lower chain can be interpreted analogously.
Unfavorable measurement outcomes, that is, outcomes other than 0, would need to be
propagated through the procedure and might affect other measurement
angles, as stated above.
If one of the measurements yields no result, the state has collapsed.
In that case, the tested function $f$ can only be balanced.
Conversely, if
the state does not collapse, $f$ is constant.

Notice that the measurement-based description exposes
the structure of the two-bit case much more clearly
than the circuit description.
As was shown in Ref.~\citen{Deutsch-Jozsa-QSL},
the resource that leads to the quantum supremacy of
the Deutsch-Jozsa over
classical algorithms is not entanglement but rather the ability of the
(quantum) oracle to apply a function to a state that is an arbitrary 
superposition of basis states.
In ZX-diagram~(\ref{eq:2-qubit-ZX-algorithm}),
it is evident that
there are two strings of information processing (along the
two linear cluster states) that do not interact with one another,
i.e., have no entanglement.
Nevertheless, the Deutsch-Jozsa algorithm
requires only one
call to the oracle, in contrast
to the two calls of the function required by the
classical algorithm.
Thus, the ZX-diagram~(\ref{eq:2-qubit-ZX-algorithm})
is in accord with and elucidates the statement of
Ref.~\citen{Deutsch-Jozsa-QSL}.


\section{Three-qubit Deutsch-Jozsa}
\subsection{Circuit model}
\label{sec:circuit3}

While, in the two-qubit case, all oracles could be realized using only
single-qubit unitary gates, the same simple procedure does not apply
for more than two qubits.
In general, one needs generalized controlled gates in higher
dimensions with arbitrary control conditions, of which the
\emph{Toffoli gate} is a simple example.
We remark that
this paper was partly inspired by the fact that
it was stated in Ref.~\citen{Cluster-Deutsch-Jozsa}
that it is possible to implement a three- or
multi-qubit Deutsch-Jozsa-algorithm
using combination of simple CNOT gates between working qubits and an
auxiliary qubit.
However, as was
shown in Refs.~\citen{Quantum_Algorithms_Josephson} and
\citen{4qubit-DJ},
this is actually
not the case---one needs a more
comprehensive approach to manage the
complexity of three or more qubits.


As in the one- and two-qubit cases, we can
write out the possible functions $f$ that are either constant or balanced.
For any number of input bits,
there are two possible constant functions, all ones or all zeros.
The number of balanced functions, however, grows super-exponentially
with the number of $n$ of input bits.
In particular, there are a total of
\begin{equation}
	N_\text{balanced} = \begin{pmatrix} 2^n \\ 2^n /
          2 \end{pmatrix} = \begin{pmatrix} 2^n
          \\ 2^{n-1} \end{pmatrix} =
        \frac{\big(2^n\big)!}{\Big(\big(2^{n-1}\big)!\Big)^2}
\end{equation}
balanced binary functions on $n$ bits,
which is the number of ways of assigning
$2^n/2$ ones to the $2^n$ possible inputs of the function $f$.
For $n = 3$, this amounts to 2 constant functions and 70 balanced
functions.

A subset of the possible functions are tabulated
in Table~\ref{tab:function3}.
If we examine variant (3), which is
\begin{equation}
	f(0, 0, 0), \, \dotsc \, , f(1, 1, 1) = 0,\, 0,\, 0,\, 0,\,
        1,\, 1,\, 1,\, 1 \; ,
\end{equation}
we see that a quantum oracle for this function is easily implemented
by adding the desired phase of $-1$ if and only if the first qubit is in state 1.
If we consider variant (4) of $f$ from
Table~\ref{tab:function3}, however, we find that the
implementation is not that straightforward.
The quantum oracle for
the balanced function
\begin{equation}
	f(0, 0, 0), \, \dotsc \, , f(1, 1, 1) = 0,\, 0,\, 0,\, 1,\,
        0,\, 1,\, 1,\, 1
\end{equation}
can no longer be implemented using only single-qubit phases.
The reason for this is that adding a phase of $-1$
to the basis vectors $\ket{011}$, $\ket{101}$, $\ket{110}$ and
$\ket{111}$ only is an \emph{entangling operation}.
That is, this operation has the
ability of transforming a product state
into an entangled state and vice versa.
In contrast, the operation in the previous example is
not an entangling operation because it can
be implemented using only
a one-qubit phase manipulation.
Entangling oracles are a phenomenon that categorically separates
the case of three or more input qubits from the case of one or two
in the Deutsch-Jozsa algorithm \cite{Quantum_Algorithms_Josephson}.

\begin{table}
\begin{center}
\caption{\label{tab:function3}Constant or balanced outputs of a three-bit boolean function. Variants (1) and (2) are constant, variants (3) through (72) are balanced.}
\begin{tabular}{cccccccccc}
\hline\hline
 & \multicolumn{8}{c}{Variants} \\
 & (1) & (2) & (3) & (4) & (5) & $\cdots$ & (71) & (72) \\
 \hline
 $f(0, 0, 0)$ & 0 & 1 & 0 & 0 & 0 & $\cdots$ & 1 & 1 \\
 $f(0, 0, 1)$ & 0 & 1 & 0 & 0 & 0 & $\cdots$ & 1 & 1 \\
 $f(0, 1, 0)$ & 0 & 1 & 0 & 0 & 0 & $\cdots$ & 1 & 1 \\
 $f(0, 1, 1)$ & 0 & 1 & 0 & 1 & 1 & $\cdots$ & 0 & 1 \\
 $f(1, 0, 0)$ & 0 & 1 & 1 & 0 & 1 & $\cdots$ & 1 & 0 \\
 $f(1, 0, 1)$ & 0 & 1 & 1 & 1 & 0 & $\cdots$ & 0 & 0 \\
 $f(1, 1, 0)$ & 0 & 1 & 1 & 1 & 1 & $\cdots$ & 0 & 0 \\
 $f(1, 1, 1)$ & 0 & 1 & 1 & 1 & 1 & $\cdots$ & 0 & 0
\end{tabular}
\end{center}
\end{table}

In order to develop a systematic approach to adding
the desired relative phases depending on the control condition, we closely
follow the approach of
Refs.~\citen{DJ-test-QC} and \citen{4qubit-DJ}.
The relative phase added by the oracle can be expressed as the unitary
mapping
\begin{equation}
	U \ket{\boldsymbol{\sigma}} = \mathrm{e}^{\mathrm{i} \theta(\boldsymbol{\sigma})}
        \ket{\boldsymbol{\sigma}} \; ,
\end{equation}
where $\ket{\boldsymbol{\sigma}} = \ket{\sigma_0 \sigma_1\dotsc}$ is a basis
vector of the multidimensional Hilbert space, e.g., $\ket{010}$, and
$\theta(\boldsymbol{\sigma}) = \pi f(\boldsymbol{\sigma})$ is the applied phase, with
$\theta \in \{0, \pi\}$.
We can write $\theta$ as
\begin{equation}
	\theta\big(\boldsymbol{\sigma}\big) = \pi \, f\big(\boldsymbol{\sigma}\big) =
        \sum_{\mathbf{x}: f(\mathbf{x}) = 1} \pi \, \delta_{\boldsymbol{\sigma},\mathbf{x}}
        \; , \label{eq:deltas}
\end{equation}
where $\delta_{\boldsymbol{\sigma},\mathbf{x}}$ is $1$ if and only if
$\boldsymbol{\sigma} = \mathbf{x}$.
This means we can write $f(\boldsymbol{\sigma})$ as sum of \emph{Kronecker
deltas}.
Since each part of the sum in Eq.~(\ref{eq:deltas}) yields one
if and only if the exact control condition of $f$ is matched
and zero otherwise, we can rewrite the 
nonzero Kronecker deltas as
\begin{equation}
	\delta_{\boldsymbol{\sigma},\mathbf{x}} = \prod_{i = 0}^{n-1} \; \Big|\sigma_i
        - x_i\Big| \; .
\end{equation}

\begin{figure}
\centering
\begin{quantikz}
\qw & \gate{P(\pi/2)} & \ctrl{2} & \qw & \qw & \qw & \ctrl{2} & \qw \\
\qw & \gate{P(\pi/2)} & \qw & \ctrl{1} & \qw & \ctrl{1} & \qw & \qw \\
\qw & \gate{P(\pi/2)} & \targ{} & \targ{} & \gate{P(-\pi/2)} & \targ{} & \targ{} & \qw
\end{quantikz}
\caption{\label{fig:example_circuit} Example of an oracle circuit for the balanced
function $f$ (variant (4) in Table~\ref{tab:function3}), where $f(000), f(001), \dotsc, f(111) = 0, 0, 0, 1, 0, 1, 1, 1$.}
\end{figure}
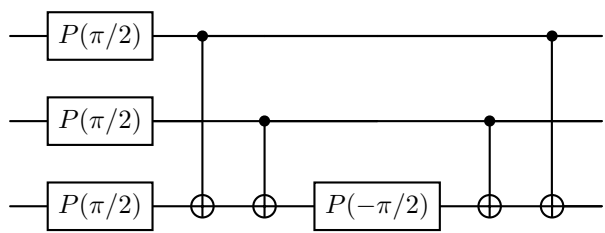

As an example, we again take function variant (4) from Table~\ref{tab:function3}.
Since the function (4) yields one for the inputs
$\mathbf{y} \in \{(011), (101), (110), (111)\}$ and zero otherwise,
we can write the non-zero deltas as
\begin{align}
	\delta_{(011),\mathbf{x}} &= (1 - x_0) x_1 x_2 = x_1 x_2 - x_0 x_1 x_2 \\
	\delta_{(101),\mathbf{x}} &= x_0(1-x_1) x_2 = x_0 x_2 - x_0 x_1 x_2 \\
	\delta_{(110),\mathbf{x}} &= x_0 x_1 (1-x_2) = x_0 x_1 - x_0 x_1 x_2 \\
	\delta_{(111),\mathbf{x}} &= x_0 x_1 x_2
\end{align}
Adding these up and using the fact that
$x_i x_j = \frac{1}{2} (x_i + x_j - x_i \oplus x_j )$ for $x_{i/j} \in \{0, 1\}$ yields 
\begin{equation}
	\begin{split}
		\theta\big(\mathbf{x}\big) &= \pi\,\Big(x_0 x_1 + x_0 x_2
                + x_1 x_2 - 2 x_0 x_1 x_2\Big) \\ 
			&= \frac{\pi}{2}\Big(x_0 + x_1 + x_2 - x_0
                \oplus x_1 \oplus x_2 \Big) \; . 
	\end{split}
        \label{eq:total_phase}
\end{equation}
The first three phases in Eq.~(\ref{eq:total_phase}) involve
non-entangling operations on the work qubits.
Namely, this can be implemented by applying a phase of $\pi/2$,
i.e., the one-qubit gate
\begin{equation}
	P(\pi/2) = \begin{pmatrix} 1 & 0 \\ 0 &
          \mathrm{e}^{\mathrm{i}\pi/2} \end{pmatrix} = \begin{pmatrix}
          1 & 0 \\ 0 & \mathrm{i} \end{pmatrix} \; , \label{eq:phase}
\end{equation}
to all three qubits.
The third term represents a phase of $-\pi/2$ if and only if there is
an odd number of ones in the basis vector.
Speaking in the language of the \emph{Clifford set}, this can be
implemented as the gate sequence
\begin{equation}
	\text{CNOT}_{01} \text{CNOT}_{12} \, P_2(-\pi/2) \,
        \text{CNOT}_{12} \text{CNOT}_{01} \; ,
\end{equation}
where $\text{CNOT}_{ij}$ represents the CNOT gate between
qubits i and j and $P_2(-\pi/2)$ represents the gate $P(-\pi/2)$
[replacing $\pi/2$ with $-\pi/2$ in
  Eq.~(\ref{eq:phase})] on qubit 2.
Putting these gates together, the oracle for function (4) from
Table~\ref{tab:function3} can be implemented as
the circuit depicted in Fig.~\ref{fig:example_circuit}.

\begin{figure}
\begin{center}
\begin{quantikz}[column sep=0.3cm]
\qw & \gate{P_1\vphantom{()}} & \ctrl{1} & \ctrl{2} & \qw & \qw & \qw & \ctrl{2} & \ctrl{1} & \qw & \qw \\
\qw & \gate{P_2\vphantom{()}} & \targ{} & \qw & \gate{P_{12}} & \ctrl{1} & \qw & \qw & \targ{} & \ctrl{1} & \qw \\
\qw & \gate{P_3\vphantom{()}} & \qw & \targ{} & \gate{P_{13}} & \targ{} & \gate{P_{123}} & \targ{} & \gate{P_{23}} & \targ{} & \qw
\end{quantikz}
\caption{\label{fig:general_circuit}General oracle circuit of the three-qubit Deutsch-Jozsa algorithm.
The one-qubit phase gates $P_\alpha$ are determined by the realization of the oracle
function $f(x_0,x_1,x_2)$
and can be constructed according to
Eqs.~(\ref{eq:P_1})--(\ref{eq:P_123}).}
\end{center}
\end{figure}
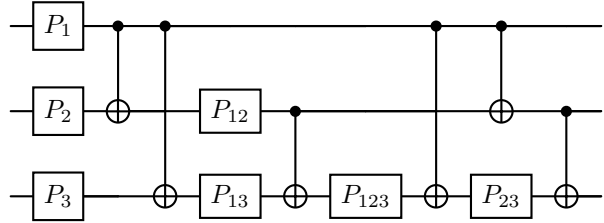

If we code the output combination of the balanced function $f$ in
binary representation, i.e., $b_7b_6\dotsc b_1b_0$,
we can express which
variant we have as
a single decimal number.
For example, variant (4) in Table~\ref{tab:function3} can be written as
$00010111_2 = 23_{10}$.
Each bit $b_i, \, i \in [0, 7]$ then corresponds to one of the eight
input basis states.
Similarly to the $P(\pi/2)$ gate,
Eq.~(\ref{eq:phase}), we can express the
general phase function as
\begin{equation}
	\begin{split}
		\theta[b_7, \dotsc, b_0]\big(\mathbf{x}\big) \\
		= \frac{\pi}{2} \bigg( 2b_7 &+ \Big(-b_6 - b_5 + b_2 +
                b_1 + \frac{1}{2} b_0 \Big) \, x_1 \\ 
			&+ \Big(-b_6 + b_4 - b_3 + b_1 + \frac{1}{2}
                b_0 \Big) \, x_2 \\ 
			&+ \Big(-b_5 + b_4 - b_3 + b_2 + \frac{1}{2}
                b_0 \Big) \, x_3 \\ 
			&- \Big(b_7 - b_5 - b_3 + b_1 + \frac{1}{2}
                b_0 \Big) \, x_1 \oplus x_2 \\ 
			&- \Big(b_7 - b_6 - b_3 + b_2 + \frac{1}{2}
                b_0 \Big) \, x_1 \oplus x_3 \\ 
			&- \Big(b_7 - b_6 - b_5 + b_4 + \frac{1}{2}
                b_0 \Big) \, x_2 \oplus x_3 \\ 
			&+ \frac{1}{2} b_0 \,  x_1 \oplus x_2 \oplus
                x_3 \bigg) \; . 
	\end{split} \label{eq:general_phase}
\end{equation}
This phase logic can be straighforwardly implemented as the circuit
depicted in Fig.~\ref{fig:general_circuit}, where
the one-qubit phase gates can be derived from
Eq.~(\ref{eq:general_phase}) as
\begin{align}
	P_1 &= P\bigg(\frac{\pi}{2} \Big(-b_6 - b_5 + b_2 + b_1 +
        \frac{1}{2} b_0 \Big) \bigg) \label{eq:P_1} \\ 
	P_2 &= P\bigg(\frac{\pi}{2} \Big(-b_6 + b_4 - b_3 + b_1 +
        \frac{1}{2} b_0 \Big) \bigg) \\ 
	P_3 &= P\bigg(\frac{\pi}{2} \Big(-b_5 + b_4 - b_3 + b_2 +
        \frac{1}{2} b_0 \Big) \bigg) \\ 
	P_{12} &= P\bigg(- \frac{\pi}{2} \Big(b_7 - b_5 - b_3 + b_1 +
        \frac{1}{2} b_0 \Big) \bigg) \\ 
	P_{13} &= P\bigg(- \frac{\pi}{2} \Big(b_7 - b_6 - b_3 + b_2 +
        \frac{1}{2} b_0 \Big) \bigg) \\ 
	P_{23} &= P\bigg(- \frac{\pi}{2} \Big(b_7 - b_6 - b_5 + b_4 +
        \frac{1}{2} b_0 \Big) \bigg) \\ 
	P_{123} &= P\bigg( \frac{\pi}{4} b_0 \bigg) \label{eq:P_123}
        \; .
\end{align}
Note that we have neglected the term $\pi\, b_7$ from
Eq.~(\ref{eq:general_phase}) because it leads solely to an overall
phase that has no physical significance.
Note also that such a circuit for the general four-qubit Deutsch-Jozsa
oracle has been derived in Ref.~\citen{4qubit-DJ}.

\subsection{Measurement-based model}
\label{sec:ZX3}

\begin{figure}
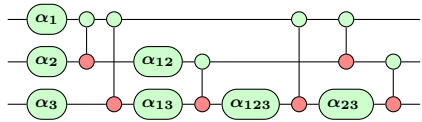

\begin{center}
\begin{ZX}
\zxN{} &[\zxwCol] \zxZ{\alpha_1} \lar \rar & \zxZ{} \dar \rar & \zxZ{} \ar[dd] \ar[rrrr] & & & & \zxZ{} \ar[dd] \rar & \zxZ{} \dar \ar[rr] & &[\zxwCol] \zxN{} \\
\zxN{} & \zxZ{\alpha_2} \lar \rar & \zxX{} \ar[rr] && \zxZ{\alpha_{12}} \rar & \zxZ{} \ar[rrr] \dar & & & \zxX{} \rar & \zxZ{} \dar \rar & \zxN{} \\
\zxN{} & \zxZ{\alpha_3} \lar \ar[rr] && \zxX{} \rar & \zxZ{\alpha_{13}} \rar & \zxX{} \rar & \zxZ{\alpha_{123}} \rar & \zxX{} \rar & \zxZ{\alpha_{23}} \rar & \zxX{} \rar & \zxN{}
\end{ZX}
\caption{\label{fig:rawZXOracle}Raw translation of the circuit in
Fig.~\ref{fig:general_circuit} to a ZX-diagram.}
\end{center}
\end{figure}

Now that we have derived the general oracle circuit for the
three-qubit Deutsch-Jozsa algorithm,
Fig.~\ref{fig:general_circuit}, we
want to translate this circuit into a ZX-diagram.
We note that we can write a CNOT operation on two qubits as 
\begin{equation}
	\begin{quantikz}[align equals at=1.5, thin lines, row sep=0.4cm]
		\qw & \ctrl{1} & \qw \\ \qw & \targ{} & \qw
	\end{quantikz} = 
	\begin{ZX}[row sep=0.4cm]
		\zxN{} &[\zxwCol] \zxZ{} \lar \rar \dar &[\zxwCol]
                \zxN{} \\
                \zxN{} & \zxX{} \lar \rar &[\zxwCol] \zxN{}
	\end{ZX} \; . \label{eq:translation_CNOT}
\end{equation}
Furthermore,
each gate in Fig.~\ref{fig:general_circuit} can be translated as
\begin{equation}
	\begin{quantikz}[align equals at= 1, thin lines]
		\qw & \gate{P(\alpha)} & \qw
	\end{quantikz} =
	\begin{ZX}
		\zxN{} &[\zxwCol] \zxZ{\alpha} \lar \rar & \zxN{} \; .
	\end{ZX} \; . \label{eq:translation_phase}
\end{equation}
Applying identities (\ref{eq:translation_CNOT}) and (\ref{eq:translation_phase}),
Fig.~\ref{fig:general_circuit} translates to the ZX-diagram depicted in
Fig.~\ref{fig:rawZXOracle}.
(A general description of the translation process and more applications
are given in Ref.~\citen{ZX-calculus}.)
This raw translation of a circuit to a ZX-diagram
is, however, not yet in a form suitable for MBQC.

In order to obtain a measurement-based description of the
three-qubit Deutsch-Jozsa algorithm, we must
simplify and rewrite the graph in Fig.~\ref{fig:rawZXOracle} so that it fulfills
the requirements for measurement-based quantum computing
outlined in Sec.~\ref{sec:ZX2}.
First, we recall that the algorithm starts with the qubits being
initialized in the $\ket{+}$ state.
It ends with the measurement of the ``all +''-state $\bra{+\dotsm+}$.
Therefore, we must add $\zx{\zxN{} &[\zxwCol] \zxZ{} \lar}$ nodes to all
incoming and outgoing legs of the diagram in Fig.~\ref{fig:rawZXOracle}
to obtain a description of the full algorithm, not just the oracle.
By successively applying rules from
the set (\ref{eq:rule_colorswap})--(\ref{eq:rule_decoupling}), we can
simplify the diagram to the one depicted in Fig.~\ref{fig:generalZXOracle}.
The simplification steps carried out are described in detail in the
Appendix.

\begin{figure}
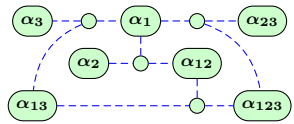

\begin{center}
\begin{ZX}
\zxZ{\alpha_3} & \zxZ{} \ar[l, blue, dashed] \ar[r, blue, dashed] \ar[ldd, blue, dashed, (] & \zxZ{\alpha_1} \ar[d, blue, dashed] \ar[r, blue, dashed] & \zxZ{} \ar[r, blue, dashed] \ar[rdd, blue, dashed, )] & \zxZ{\alpha_{23}} \\
\zxN{} & \zxZ{\alpha_2} \ar[r, blue, dashed] & \zxZ{} \ar[r, blue, dashed] & \zxZ{\alpha_{12}} \ar[d, blue, dashed] & \zxN{} \\
\zxZ{\alpha_{13}} \ar[rrr, blue, dashed] &&& \zxZ{} \ar[r, blue, dashed] & \zxZ{\alpha_{123}}
\end{ZX}
\caption{\label{fig:generalZXOracle}
  Full ZX-diagram of the three-qubit Deutsch-Jozsa algorithm in
  measurement form.}
\end{center}
\end{figure}

The diagram of Fig.~\ref{fig:generalZXOracle}
can be interpreted in the language of MBQC as follows:
Prepare a cluster state with the geometry of the diagram, where nodes
are qubits and edges depict entanglement.
Then measure all qubits in their respective bases at the given angles.
If no angle is given, measure in the $x$-basis.
The angles of measurement to implement a given realization of the
oracle are given by Eqs.~(\ref{eq:P_1})--(\ref{eq:P_123}).
For unfavorable measurement outcomes,
that is, measurements that do not yield the first outcome,
corrective measurement angles are
propagated through the diagram, as mentioned above and
explained in detail in Ref.~\citen{Circuit-extraction-tale}.
Since we have derived a graph with a measurement pattern
whose outputs are all in the $\ket{+}$-state, the following statement holds:
If any of the measurements yield no outcome, meaning that the state has
collapsed beforehand, then the tested function is balanced.
If we obtain an outcome for all measurements,
then the tested function is constant.

\begin{figure*}
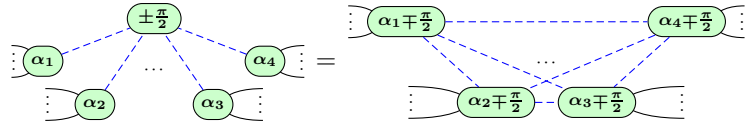

	\[\begin{ZX}
		\zxN{} &&& \zxZ{\pm\frac{\pi}{2}} \ar[dd, 3 dots] && \zxN{} \\
		\leftManyDots{} \zxZ{\alpha_1} \ar[rru, blue, dashed] &&&& \zxZ{\alpha_4} \ar[llu, blue, dashed] \rightManyDots{} \\
		& \leftManyDots{} \zxZ{\alpha_2} \ar[ruu, blue, dashed] & \zxN{} & \zxZ{\alpha_3} \ar[luu, blue, dashed] \rightManyDots{} &
	\end{ZX} =
	\begin{ZX}
		\leftManyDots{} \zxZ{\alpha_1\mp\frac{\pi}{2}} \ar[rd, blue, dashed] \ar[rrrr, blue, dashed] \ar[rrrd, blue, dashed] & & \zxN{} \ar[d, 3 dots] & & \zxZ{\alpha_4\mp\frac{\pi}{2}} \ar[ld, blue, dashed] \ar[llld, blue, dashed] \rightManyDots{} \\[0.5cm]
		& \leftManyDots{} \zxZ{\alpha_2\mp\frac{\pi}{2}} \ar[rr, blue, dashed] & \zxN{} & \zxZ{\alpha_3\mp\frac{\pi}{2}} \rightManyDots{}
	\end{ZX}\]
\caption{\label{fig:rule_bridging}Complementing rule of ZX-calculus. A $\pi/2$ measurement in
the XY-plane leads to complementing the subgraph consisting
of all qubits in the neighborhood of the measured qubit
plus a phase correction. The measured qubit is removed from
the graph.}
\end{figure*}

\begin{figure}
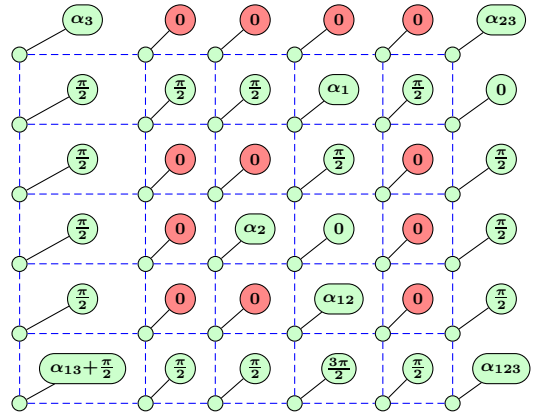

\begin{center}
\begin{ZX}
\zxN{} & \zxZ{\alpha_3} && \zxX{0} && \zxX{0} && \zxX{0} && \zxX{0} && \zxZ{\alpha_{23}} \\
\zxZ{} \ar[ur] \ar[dd, blue, dashed] \ar[rr, blue, dashed] && \zxZ{} \ar[ur] \ar[dd, blue, dashed] \ar[rr, blue, dashed] && \zxZ{} \ar[ur] \ar[dd, blue, dashed] \ar[rr, blue, dashed] && \zxZ{} \ar[ur] \ar[dd, blue, dashed] \ar[rr, blue, dashed] && \zxZ{} \ar[ur] \ar[dd, blue, dashed] \ar[rr, blue, dashed] && \zxZ{} \ar[ur] \ar[dd, blue, dashed] & \\
\zxN{} & \zxFracZ{\pi}{2} && \zxFracZ{\pi}{2} && \zxFracZ{\pi}{2} && \zxZ{\alpha_1} && \zxFracZ{\pi}{2} && \zxZ{0} \\
\zxZ{} \ar[ur] \ar[dd, blue, dashed] \ar[rr, blue, dashed] && \zxZ{} \ar[ur] \ar[dd, blue, dashed] \ar[rr, blue, dashed] && \zxZ{} \ar[ur] \ar[dd, blue, dashed] \ar[rr, blue, dashed] && \zxZ{} \ar[ur] \ar[dd, blue, dashed] \ar[rr, blue, dashed] && \zxZ{} \ar[ur] \ar[dd, blue, dashed] \ar[rr, blue, dashed] && \zxZ{} \ar[ur] \ar[dd, blue, dashed] & \\
\zxN{} & \zxFracZ{\pi}{2} && \zxX{0} && \zxX{0} && \zxFracZ{\pi}{2} && \zxX{0} && \zxFracZ{\pi}{2} \\
\zxZ{} \ar[ur] \ar[dd, blue, dashed] \ar[rr, blue, dashed] && \zxZ{} \ar[ur] \ar[dd, blue, dashed] \ar[rr, blue, dashed] && \zxZ{} \ar[ur] \ar[dd, blue, dashed] \ar[rr, blue, dashed] && \zxZ{} \ar[ur] \ar[dd, blue, dashed] \ar[rr, blue, dashed] && \zxZ{} \ar[ur] \ar[dd, blue, dashed] \ar[rr, blue, dashed] && \zxZ{} \ar[ur] \ar[dd, blue, dashed] & \\
\zxN{} & \zxFracZ{\pi}{2} && \zxX{0} && \zxZ{\alpha_2} && \zxZ{0} && \zxX{0} && \zxFracZ{\pi}{2} \\
\zxZ{} \ar[ur] \ar[dd, blue, dashed] \ar[rr, blue, dashed] && \zxZ{} \ar[ur] \ar[dd, blue, dashed] \ar[rr, blue, dashed] && \zxZ{} \ar[ur] \ar[dd, blue, dashed] \ar[rr, blue, dashed] && \zxZ{} \ar[ur] \ar[dd, blue, dashed] \ar[rr, blue, dashed] && \zxZ{} \ar[ur] \ar[dd, blue, dashed] \ar[rr, blue, dashed] && \zxZ{} \ar[ur] \ar[dd, blue, dashed] & \\
\zxN{} & \zxFracZ{\pi}{2} && \zxX{0} && \zxX{0} && \zxZ{\alpha_{12}} && \zxX{0} && \zxFracZ{\pi}{2} \\
\zxZ{} \ar[ur] \ar[dd, blue, dashed] \ar[rr, blue, dashed] && \zxZ{} \ar[ur] \ar[dd, blue, dashed] \ar[rr, blue, dashed] && \zxZ{} \ar[ur] \ar[dd, blue, dashed] \ar[rr, blue, dashed] && \zxZ{} \ar[ur] \ar[dd, blue, dashed] \ar[rr, blue, dashed] && \zxZ{} \ar[ur] \ar[dd, blue, dashed] \ar[rr, blue, dashed] && \zxZ{} \ar[ur] \ar[dd, blue, dashed] & \\
\zxN{} & \zxZ{\alpha_{13}+\frac{\pi}{2}} && \zxFracZ{\pi}{2} && \zxFracZ{\pi}{2} && \zxFracZ{3\pi}{2} && \zxFracZ{\pi}{2} && \zxZ{\alpha_{123}} \\
\zxZ{} \ar[ur] \ar[rr, blue, dashed] && \zxZ{} \ar[ur] \ar[rr, blue, dashed] && \zxZ{} \ar[ur] \ar[rr, blue, dashed] && \zxZ{} \ar[ur] \ar[rr, blue, dashed] && \zxZ{} \ar[ur] \ar[rr, blue, dashed] && \zxZ{} \ar[ur] &
\end{ZX}
\caption{\label{fig:lattice}Lattice version of the cluster-state Deutsch-Jozsa
algorithm. The pattern in Fig.~\ref{fig:generalZXOracle} is
embedded into this lattice. Spare qubits are removed or
bridged with X-measurements or $\pi/2$-XY-measurements.}
\end{center}
\end{figure}

For experimental implementations, it can be convenient
to embed the aforementioned
pattern into a rectangular lattice cluster state because such lattice states
are more straightforward to treat
than cluster states corresponding to arbitrary graphs,
i.e., do not require special attention to the geometry of the resource state.
By adding a few more qubits, we can extract the
pattern of our algorithm from a rectangular lattice.
We can use rule (\ref{eq:rule_decoupling}) to
decouple qubits from a cluster state and thus effectively remove them.
A more advanced rule of ZX-calculus, depicted in Fig.~\ref{fig:rule_bridging},
states that a measurement at an angle of $\pm\pi/2$ in the XY-plane
leads to the removal of that qubit along with complementing
the subgraph consisting of all qubits in the neighborhood of the qubit.
Complementation of a
neighborhood means that all edges connecting
the nodes inside the set defining the neighborhood are removed and
that all edges that formerly were not present are added.
Thus, the edge allocation among the distribution of all possible edges
between the nodes connected to the measured qubit is inverted.
For a more thorough description, see Ref.~\citen{ZX-simplification}.
Given these two rules, it is straightforward to show that the diagram
of Fig.~\ref{fig:lattice} is equivalent to that of Fig.~\ref{fig:generalZXOracle}.
We state explicitly that we do not claim to have proven this lattice to be
the smallest possible lattice cluster state that can contain
the three-qubit Deutsch-Jozsa algorithm.
Instead, our formulation should be regarded as a proof of concept.

\section{Summary and Discussion}
\label{sec:Conclusion}

We have used the
ZX-calculus to
reformulate the two- and three-qubit Deutsch-Jozsa algorithms
in the language of MBQC,
translating the circuit descriptions to MBQC-compatible ZX-diagrams.
The result is a pattern that describes both the geometry of the
underlying cluster state as well as a sequence of single-qubit
measurements that must be carried out in order to implement a
particular realization of the oracle.
In addition, we have derived an alternate, but equivalent, ZX-diagram that 
has rectangular form.
This variant could potentially be useful for experimental implementations.

For the two-qubit algorithm, we have found a comprehensible
illustration for the thesis of
Ref.~\citen{Deutsch-Jozsa-QSL} that the quantum supremacy
in the Deutsch-Jozsa algorithm lies in the ability of the
quantum oracle to operate on a superposition state, i.e., to utilize
quantum parallelism, rather than in any use of entanglement.
This illustrates that the MBQC formulation of an algorithm, coupled
with its representation in terms of ZX-calculus, can elucidate the
underlying structure of a quantum computation and clarify the extent
and origin of the quantum advantage.
We note that there are other routes to such a clarification, for
example, via the topological
description of quantum algorithms \cite{Topological_Algorithms}.
Both the topological description and the ZX-calculus itself can be
viewed as being part of the more fundamental and general framework
of \emph{categorical quantum mechanics}; 
for more information, we refer the interested reader to
Refs.~\citen{CQM1} and \citen{CQM2}.
An MBQC implementation of the
two-qubit version of the Deutsch-Josza algorithm has already been
formulated in Ref.~\citen{Cluster-Deutsch-Jozsa}.
However, the authors did not include a derivation of their MBQC formulation.
We remark that the two-qubit case is sufficiently simple so that educated guessing
is sufficient to find a working and presumably optimal MBQC implementation;
in our opinion, this is no longer possible for the three-qubit case.
Ref.~\citen{Cluster-Deutsch-Jozsa}
also includes a generalization to $n$ qubits
that is based on the assumption that the two- and more-qubit versions
of the algorithm have
the same structural complexity.
However, it was shown in Ref.~\citen{Quantum_Algorithms_Josephson}
that this is not the case, and our work, based on a different,
graphical approach, also contradicts this simplified generalization.

We hope that the measurement-based formulation of the three-qubit
version of the Deutsch-Jozsa algorithm that we
have described in this article can serve as a blueprint
for an experimental implementation as well as a step on the way to
formulating an MBQC description of the $n$-qubit algorithm.

While the translation from quantum circuit to
ZX-representation
is well-known, and the conversion to MBQC-form is
relatively straightforward,
we know of no explicit reformulations of quantum algorithms using
this method in the previous literature.
In any case,
subsequent simplification of the ZX-diagram is less
straightforward, at least if an optimal pattern that minimizes the size of the
cluster state is the goal.
This question of optimal minimization cannot be answered
through the framework of ZX-calculus because ZX-calculus
is only a descriptive language.
However, the rules of ZX-calculus could be a good tool to visualize a
path to a minimal solution, as is discussed
in Ref.~\citen{ZX-simplification}.
Thus, interesting subjects for future work would be 
to investigate if the patterns derived here do, in fact, yield a minimal
cluster state and, in addition,
to formulate schemes for finding such
patterns in general and for proving that they are, in fact, minimal.

While we have specifically treated the Deutsch-Jozsa algorithm here,
the methods we have outlined are relatively general and can be applied
to other quantum circuits, i.e., to other quantum algorithms.
Thus, our treatment here is intended to illustratrate the scheme using
a simple but still nontrivial example.
An obvious extension would be to Deutsch-Jozsa algorithms with more
than three qubits; a general scheme for the $n$-qubit case would be
particularly interesting.
Applying the method to other quantum algorithms in general, and
to those that are based on a mutable oracle in particular, would be a 
an important longer-term goal;
the Shor factorization algorithm is a particularly  prominent
example of such an algorithm.

\appendix*

\section{Simplification of the oracle in the ZX-language}
\label{app:simplification}

In this appendix, we describe the detailed translation and
simplification steps used to reformulate
the general three-qubit Deutsch-Jozsa algorithm in terms of MBQC.
Starting with the circuit of Fig.~\ref{fig:general_circuit},
we use the identities
\begin{equation}
	\begin{quantikz}[align equals at=1.5, thin lines, row sep=0.4cm]
		\qw & \ctrl{1} & \qw \\ \qw & \targ{} & \qw
	\end{quantikz} = 
	\begin{ZX}[row sep=0.4cm]
		\zxN{} &[\zxwCol] \zxZ{} \lar \rar \dar &[\zxwCol]
                \zxN{} \\
                \zxN{} & \zxX{} \lar \rar &[\zxwCol] \zxN{}
	\end{ZX}
	\quad \text{and} \quad 
	\begin{quantikz}[align equals at= 1, thin lines]
		\qw & \gate{P(\alpha)} & \qw
	\end{quantikz} =
	\begin{ZX}
		\zxN{} &[\zxwCol] \zxZ{\alpha} \lar \rar & \zxN{} \; .
	\end{ZX}
\end{equation}
to derive the raw translation depicted in Fig.~\ref{fig:rawZXOracle}.
We then add one $\zx{\zxN{} &[\zxwCol] \zxZ{} \lar}$ node to the leg
protruding to the right to represent the initial state $\ket{+++}$
and  another to the left to represent the measurement of
the same state, i.e., multiply from the left with $\bra{+++}$.
This leaves us with the closed ZX-diagram
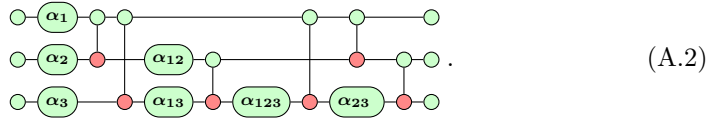
\begin{equation}
\begin{ZX}
\zxZ{} & \zxZ{\alpha_1} \lar \rar & \zxZ{} \dar \rar & \zxZ{} \ar[dd] \ar[rrrr] & & & & \zxZ{} \ar[dd] \rar & \zxZ{} \dar \ar[rr] & & \zxZ{} \\
\zxZ{} & \zxZ{\alpha_2} \lar \rar & \zxX{} \ar[rr] && \zxZ{\alpha_{12}} \rar & \zxZ{} \ar[rrr] \dar & & & \zxX{} \rar & \zxZ{} \dar \rar & \zxZ{} \\
\zxZ{} & \zxZ{\alpha_3} \lar \ar[rr] && \zxX{} \rar & \zxZ{\alpha_{13}} \rar & \zxX{} \rar & \zxZ{\alpha_{123}} \rar & \zxX{} \rar & \zxZ{\alpha_{23}} \rar & \zxX{} \rar & \zxZ{}
\end{ZX} \; .
\label{eq:rawZXdiagram}
\end{equation}
Since the diagram has no protruding legs, it represents a scalar.
Thus, it immediately determines the result of all measurements.
In ZX-diagram (\ref{eq:rawZXdiagram}),
we apply rule~(\ref{eq:rule_colorswap})
to all X-nodes (red) and rule~(\ref{eq:rule_contraction})
to all adjacent Z-nodes (green) that are connected through regular
edges (solid), arriving at the diagram
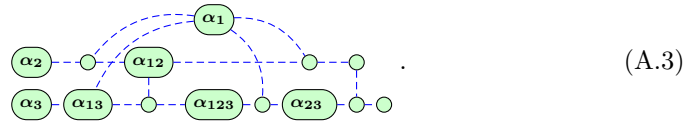
\begin{equation}
\begin{ZX}
\zxN{} &&& \zxZ{\alpha_1} \ar[lld, blue, dashed, (] \ar[lldd, blue, dashed, (] \ar[rdd, blue, dashed, )] \ar[rrd, blue, dashed, )] &&&& \zxN{} \\
\zxZ{\alpha_2} \ar[r, blue, dashed] & \zxZ{} \ar[r, blue, dashed] & \zxZ{\alpha_{12}} \ar[rrr, blue, dashed] \ar[d, blue, dashed] &&& \zxZ{} \ar[r, blue, dashed] & \zxZ{} \ar[d, blue, dashed] & \zxN{} \\
\zxZ{\alpha_3} \ar[r, blue, dashed] & \zxZ{\alpha_{13}} \ar[r, blue, dashed] & \zxZ{} \ar[r, blue, dashed] & \zxZ{\alpha_{123}} \ar[r, blue, dashed] & \zxZ{} \ar[r, blue, dashed] & \zxZ{\alpha_{23}} \ar[r, blue, dashed] & \zxZ{} \ar[r, blue, dashed] & \zxZ{}
\end{ZX} \; .
\end{equation}
We notice that we can remove the rightmost node on the middle row
because $\zx{\zxN{} &[\zxwCol] \zxZ{} \ar[l, blue, dashed] \ar[r,
    blue, dashed] &[\zxwCol] \zxN{}} = \zx{\zxN{} \rar &[\zxWCol]
  \zxN{}}$ ($H^2 = I$).
Furthermore, we again use rule (\ref{eq:rule_colorswap})
to change the ``tail'' on the
rightmost end of the third row to an X-node (red).
This yields the diagram
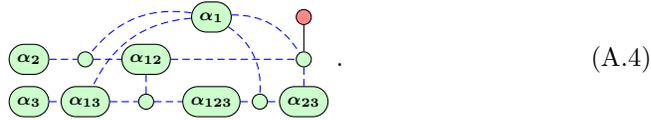
\begin{equation}
	 \begin{ZX}
		\zxN{} &&& \zxZ{\alpha_1} \ar[lld, blue, dashed, (] \ar[lldd, blue, dashed, (] \ar[rdd, blue, dashed, )] \ar[rrd, blue, dashed, )] && \zxX{} \dar \\
		\zxZ{\alpha_2} \ar[r, blue, dashed] & \zxZ{} \ar[r, blue, dashed] & \zxZ{\alpha_{12}} \ar[rrr, blue, dashed] \ar[d, blue, dashed] &&& \zxZ{} \ar[d, blue, dashed] \\
		\zxZ{\alpha_3} \ar[r, blue, dashed] & \zxZ{\alpha_{13}} \ar[r, blue, dashed] & \zxZ{} \ar[r, blue, dashed] & \zxZ{\alpha_{123}} \ar[r, blue, dashed] & \zxZ{} \ar[r, blue, dashed] & \zxZ{\alpha_{23}}
	\end{ZX} \; .
\end{equation}
We now apply the rule
\begin{equation}
	\begin{ZX}
		\zxN{} & &[\zxwCol] \zxN{}\\[\zxwRow]
		\zxX{} \rar & \zxZ{\alpha} \ar[ru, )] \ar[rd, (] \ar[r, 3 vdots] & \zxN{} \\[\zxwRow]
		& & \zxN{}
	\end{ZX}
	= 
	\begin{ZX}
		\zxX{} \rar &[\zxwCol] \zxN{} \\[\zxwRow]
		\zxN{} \ar[r, 3 vdots] & \zxN{} \\[\zxwRow]
		\zxX{} \rar & \zxN{}
	\end{ZX} \; ,
\end{equation}
which states that an X-measurement (X-node/red node) on a Z-node
(green) will decouple and erase the latter inside the graph,
to the rightmost node in the middle row, obtaining
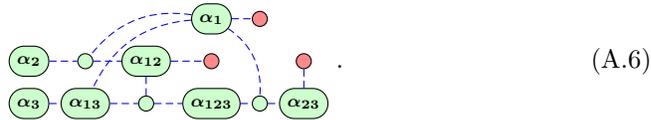
\begin{equation}
\begin{ZX}
\zxN{} &&& \zxZ{\alpha_1} \ar[lld, blue, dashed, (] \ar[lldd, blue, dashed, (] \ar[rdd, blue, dashed, )] \ar[r, blue, dashed] & \zxX{} & \zxN{} \\
\zxZ{\alpha_2} \ar[r, blue, dashed] & \zxZ{} \ar[r, blue, dashed] & \zxZ{\alpha_{12}} \ar[r, blue, dashed] \ar[d, blue, dashed] & \zxX{} && \zxX{} \ar[d, blue, dashed] \\
\zxZ{\alpha_3} \ar[r, blue, dashed] & \zxZ{\alpha_{13}} \ar[r, blue, dashed] & \zxZ{} \ar[r, blue, dashed] & \zxZ{\alpha_{123}} \ar[r, blue, dashed] & \zxZ{} \ar[r, blue, dashed] & \zxZ{\alpha_{23}}
\end{ZX} \; .
\end{equation}
Finally, we notice that $\zx{\zxN{} &[\zxwCol] \zxX{} \ar[l, blue,
    dashed]} = \zx{\zxN{} &[\zxwCol] \zxZ{} \lar}$ because of
rule~(\ref{eq:rule_colorswap}).
This means that, using rule
(\ref{eq:rule_contraction}), we can merge
the X-node (red) into its neighbor.
After some rearrangement, we obtain
the final result in Fig.~(\ref{fig:generalZXOracle}).


\end{document}